\documentclass[prx,twocolumn,showpacs,superscriptaddress]{revtex4-1}
 
\usepackage{multirow,eurosym,amssymb,amsfonts,amsmath,setspace,graphicx,color,bm,float,verbatim}

\newcommand{\bra}[1]{\langle #1|}
\newcommand{\ket}[1]{|#1\rangle}

\def\ket#1{| #1 \rangle}
\def\bra#1{\langle #1 |}

\def\be{\begin{equation}}
\def\ee{\end{equation}}

\def\bsplit{\begin{split}}
\def\nsplit{\end{split}}

\begin{document}
\title{Implementation of traveling odd Schr\"odinger cat states in circuit-QED}

\date{\today}
\author{Jaewoo Joo}
\affiliation{School of Computational Sciences, Korea Institute for Advanced Study, Seoul 02455, Korea}

\author{Su-Yong Lee} 
\affiliation{School of Computational Sciences, Korea Institute for Advanced Study, Seoul 02455, Korea}
  
\author{Jaewan Kim} 
\affiliation{School of Computational Sciences, Korea Institute for Advanced Study, Seoul 02455, Korea}

\begin{abstract}
We propose a realistic scheme of generating a traveling odd Schr\"odinger cat state and a generalized entangled coherent state in circuit quantum electrodynamics (circuit-QED). A squeezed vacuum state is used as initial resource of nonclassical states, which can be created through a Josephson traveling-wave parametric amplifier, and travels through a transmission line. Because a single-photon subtraction from the squeezed vacuum gives with very high fidelity an odd Schr\"odinger cat state, we consider a specific circuit-QED setup consisting of the Josephson amplifier creating the traveling resource in a line, a beam-splitter coupling two transmission lines, and a single photon detector located at the end of the other line. When a single microwave photon is detected by measuring the excited state of a superconducting qubit in the detector, a heralded cat state is generated with high fidelity in the opposite line. For example, we show that the high fidelity of the outcome with the ideal cat state can be achieved with appropriate squeezing parameters theoretically. As its extended setup, we suggest that generalized entangled coherent states can be also built probabilistically and useful for microwave quantum information processing for error-correctable qudits in circuit-QED.
\end{abstract}

\maketitle
\begin{figure}[b]
\includegraphics[width=8.5cm,trim= 1cm 1cm 1cm  1cm]{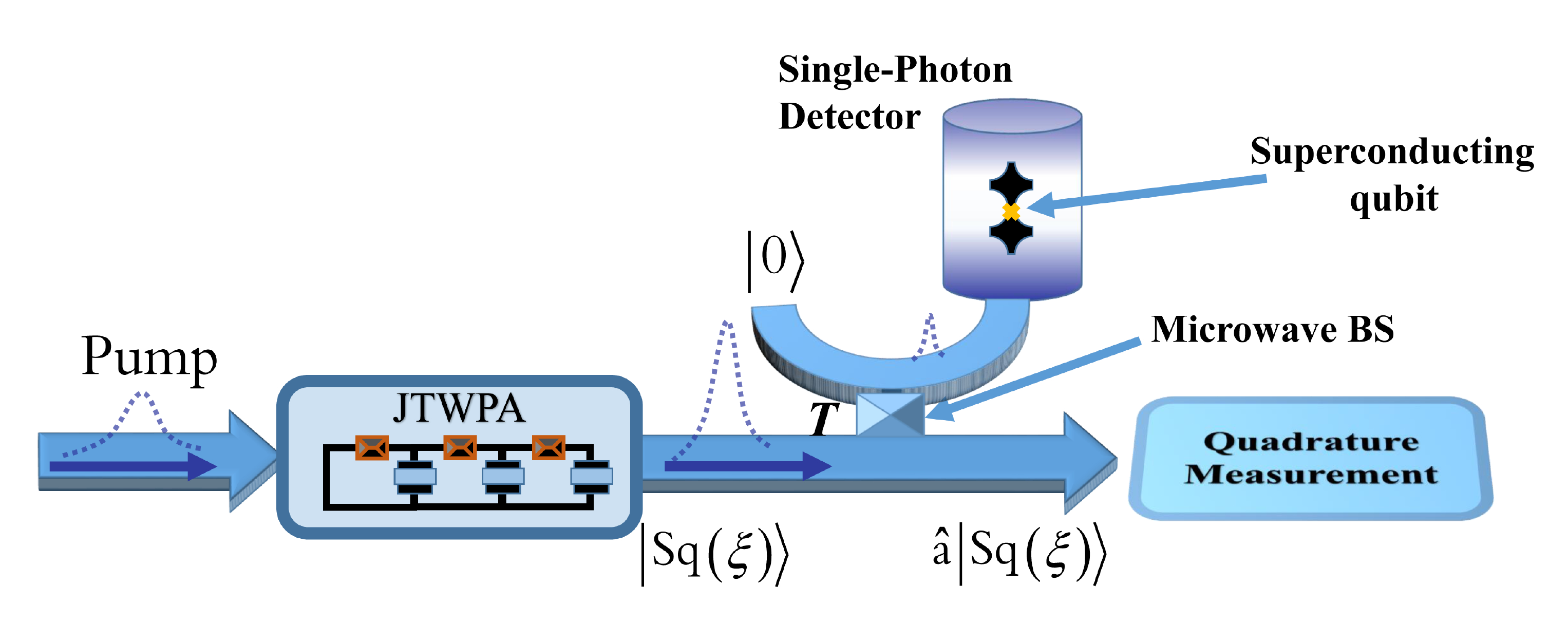}
\caption{Schematics for generating approximate traveling cat states in circuit-QED. After the JTWPA, a squeezed vacuum state $\ket{Sq(\xi)}$ is propagating in a transmission line. A microwave beam splitter (BS) with transmission $T$ is located between two transmission lines. Once a microwave photon is detected using a superconducting qubit inside a cavity, the state becomes a single-photon subtracted SVS in the other line approximately. This can be verified by quadrature measurements.}
\label{fig:01}
\end{figure}

\section{Background}
Since the thought-experiment of Schr\"odinger cat states (SCSs) was proposed as an example of macroscopic superposed states in 1935 \cite{Schro35},
the implementation of this quantum coherence has been investigated in various physical systems. One of the well-known representations for the SCS is the superposition of two coherent states with opposite phases \cite{Yurke86} and the research on this nonclassical state has great potential not only for understanding fundamental quantum physics but also for openning up new avenues for quantum technologies (e.g., continuous-variable quantum computing \cite{CVQC}, quantum metrology \cite{ JooPRL11,Su-Yong,Su-Yong1}, and quantum communications \cite{CVQKD}). Quantum optics has provided an excellent platform for generating both trapped and traveling SCSs. For example, the stationary SCS has been firstly demonstrated in an optical cavity system interacted with flying atoms \cite{Nobel12,NewHaroche08}. 

We here focus on how to generate the traveling SCSs, which could be used as a carrier of quantum information (QI) from one QI processing unit to the other. An optical parametric amplifier enables to create nonclassical photons in the setup of quantum optics  (e.g., a squeezed vacuum state (SVS)) \cite{photon-cat}. In particular, the scheme of subtracting photons has been theoretically proposed to achieve the approximated SCSs with high fidelity \cite{Jeong_MSK_PRA09} and been experimentally realized with the moderate size of its amplitude in the optical SCSs \cite{Exp_SSV,Jeong07}. It is known that the SCSs with a sufficiently large amplitude are utilized for fault-tolerant continuous-variable QI processing \cite{Ralph08}. Because this method brings the very high fidelity to SCSs at small amplitudes, the amplification of the SCSs may be required for practical QI processing \cite{amplification}. In addition, the traveling SCSs are also useful resource states for performing nonlocality tests \cite{Sanders95, Jeong08} and for testing indistinguishability of macroscopic states \cite{Su-Yong2}.

The rapid development of superconducting circuit technology has shown the potential to provide a new platform for scalable quantum systems. The SCS has been recently built inside a 3D microwave cavity coupled with a specific superconducting qubit and quantum Zeno dynamics has been also used for implementing SCSs \cite{Yale_big_cat}. A dispersive Hamiltonian brings the capability to control cavity states useful for continuous-variable QI processing in a cavity-superconductor architecture, and entangled coherent states have been very recently created inside two cavities jointed with a superconducting qubit \cite{Yale-entangle}. 

In particular, the Josephson junction non-linearity has been recently used for realizing the Josephson parametric amplifier (JPA) \cite{JPA} or Josephson traveling-wave parametric amplifier (JTWPA) to generate traveling nonclassical microwave photons in circuit-quantum electrodynamics (circuit-QED) \cite{JTWPA}. We here do not distinguish between the JPA and JTWPA because both produce a traveling SVS even though they are quite different from each other on other characteristics. Using a traveling photon resource from JTWPA/JPA, many interesting experiments have performed in circuit-QED beyond conventional experiments in quantum optics \cite{Siddiqi,Nakamura_PRL} and this technical development allows us to investigate traveling microwave qubits through transmission lines corresponding to a photonic QI processing in quantum optics \cite{GrangierECS}. For example, the most recent experiments have shown the detection schemes of a single microwave photon with excellent detection efficiency \cite{1microwaveD}, which will be a key ingredient for traveling-microwave QI processing.

We here propose a scheme of generating the odd SCSs propagating in a transmission line in circuit-QED. For its implementation, a JTWPA and a single microwave detector are basically required in superconducting circuits. In Fig.~\ref{fig:01}, a SVS $\ket{Sq}$ is traveling in the 1D transmission line with which the other line is coupled by a microwave beam-splitter (BS). If the detector of a single microwave photon is located in the other line, the event of the photon measurement reveals the single-photon subtraction from the SVS and the output state has very high fidelity compared with the traveling odd SCS. Finally, we show that the extension of this idea can be applied for creating traveling generalized entangled cat states, which may be useful for qudit QI processing and microwave logical qubits in quantum error-correction \cite{Yale-QECC}.

\begin{figure}[t]
\includegraphics[width=8.5cm,trim= 0cm 1cm 0cm  0cm]{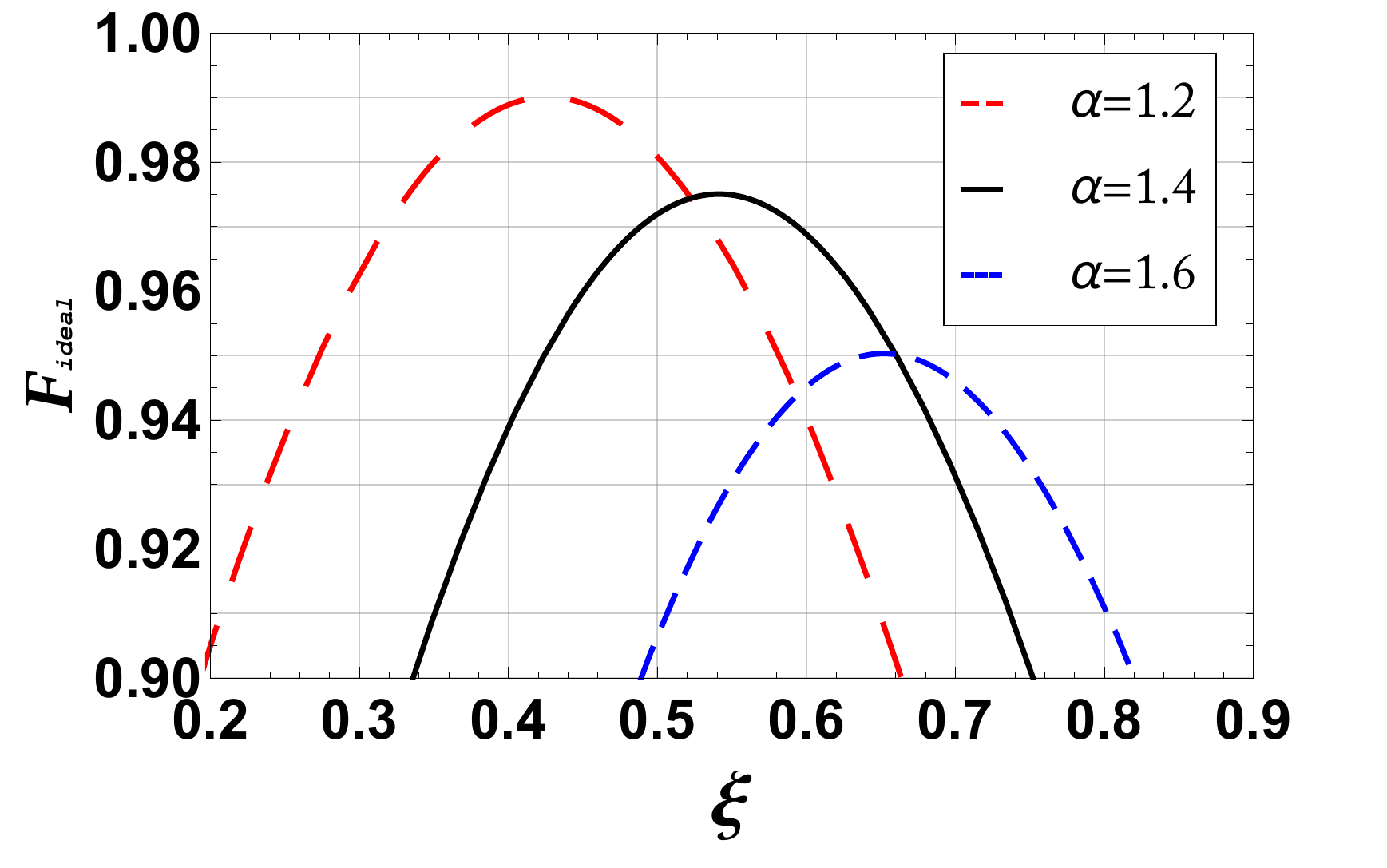}
\caption{Fidelity $F_{ideal}$ between an ideal single-photon subtracted SVS $\ket{\psi_{3Sq} (\xi)} $ and an odd SCS $\ket{SCS^-_{i \alpha}}$ for $\alpha =$ 1.2, 1.4, and 1.6.}
\label{fig:02}
\end{figure}

\section{Photon-subtraction from a SVS}
An even/odd Schr\"odinger cat state is presented by 
\begin{eqnarray}
\label{SCS01}
\ket{SCS^\pm_{\beta}} =  M^\pm_{\beta} (\ket{\beta} \pm \ket{-\beta}),~~~~~
\end{eqnarray}
where $M^{\pm}_{\beta} = 1/\sqrt{2(1\pm e^{-2|\beta|^2})}$ and the coherent state is 
\begin{eqnarray}
\label{SCS02}
\ket{\beta} =  e^{-|\beta|^2 \over 2} \sum_{m=0}^{\infty} {\beta^m \over \sqrt{m!}} \, \ket{m}. 
\label{Coherent01}
\end{eqnarray}
Note that $\bra{-\beta} \beta \rangle = e^{-2|\beta|^2} \approx 0$ and $M^\pm_{\beta} \approx 1/\sqrt{2}$ for bigger $\beta$.
In particular, the odd SCS $\ket{SCS^-_{\beta}} $ can be reformulated in the superposition of odd photon-number states ($m=2n+1$),
\begin{eqnarray}
\label{SCS03}
\ket{SCS^-_{\beta}} =  2 M^-_{\beta}  e^{-|\beta|^2 \over 2} \sum_{n=0}^{\infty} {\beta^{2n+1} \over \sqrt{(2n+1)!}} \, \ket{2n+1}.~~~~~
\end{eqnarray}

It is known that an odd SCS can be approximately generated from a SVS with a single-photon subtraction and we here assume that the single-photon subtraction operation is ideally represented by the photon annihilation operation $\hat{a}$. The SVS is given by
\begin{eqnarray}
&& \ket{Sq (\xi) } = \sqrt{{\rm sech\,} \xi} \sum_{l=0}^{\infty} {\sqrt{(2l)!} \over l! } \left( - {1 \over 2} {\rm tanh}\, \xi \right)^l \ket{2l},~~~ \label{Sq_V01}
\end{eqnarray}
which contains only even photon-number states ($\xi$; squeezing parameter)\cite{Jeong_MSK_PRA09}. 
Then, if a single photon is subtracted from $\ket{Sq (\xi) }$, the output state (named single-photon subtracted SVS $\ket{\psi_{Sq}}$) is given by
\begin{eqnarray}
\label{Sq_V02}
\ket{\psi_{Sq} (\xi)} = {1 \over {\rm sinh} \, \xi} \, \hat{a}\, \ket{Sq (\xi) },
\end{eqnarray}
which is normalized by the factor $1/ {\rm sinh} \, \xi$.
Note that this state only contains odd photon number states.
Thus, the fidelity between $\ket{\psi_{Sq}}$ and $\ket{SCS^-_{\beta=i\alpha}}$ \cite{CV_fidelity} is given by 
\begin{eqnarray}
\label{Fidelity01}
 F_{ideal} &&= |\bra{\psi_{Sq} (\xi)}  SCS^-_{i \alpha} \rangle |^2. ~~~
\end{eqnarray}
In Fig.~\ref{fig:02}, $\ket{\psi_{Sq} (\xi)}$ reaches to an odd Schr\"odinger cat state with the maximum fidelity as a function of $\xi$ and $\alpha$ and a high fidelity can be achieved with the condition of $F \ge 0.95$ for $\alpha \le 1.6$.
For example, $F \approx 0.99$ at $\alpha=1.2$ and $\xi = 0.43$, $F \approx 0.975$ at $\alpha=1.4$ and $\xi = 0.54$ and $F \approx 0.95$ at $\alpha=1.6$ and $\xi = 0.65$ as shown in Fig.~\ref{fig:02}.

\section{Implementation in circuit-QED}
\subsection{Protocol}
We present a novel scheme for implementation of a traveling Schr\"odinger cat state from a SVS produced through the JTWPA. The proposed scheme has two main parts in Fig.~\ref{fig:01}. The first part before a microwave BS shows how to generate the traveling SVS, which has been recently implemented in several world-leading groups \cite{JTWPA, Siddiqi, Nakamura_PRL}. The JTWPA is formed in a chain of Josephson junctions, capacitors, and inductors and the technique of resonant phase matching is used in a four-wave mixing process either to amplify an input signal or to generate a SVS. The bandwidth of the JTWPA is about some GHz and the four-wave mixing process brings the squeezing of the output microwave at the end of the JTWPA. Thus, the squeezed state from the JTWPA might be used as an important resource of traveling microwave states for QI processing in circuit-QED.

As shown in Fig.~\ref{fig:01}, the protocol is as follows. 1) A SVS $\ket{Sq (\xi) }_A$ is generated by the JTWPA in mode $A$. 2)  $\ket{Sq (\xi) }$ passes through a transmission line and a microwave BS makes a coupling with transmission $T$ between the transmission line and an additional line $A'$. 3) At the end of the line $A'$, a superconducting qubit is located to detect a single-photon microwave in a cavity. 4) If the detector clicks, a single-photon subtracted SVS is heralded in mode $A$ and a quadrature measurement is performed to verify the odd SCSs.
\begin{figure}[t]
\includegraphics[width=8.5cm,trim= 0cm 0cm 0cm  0cm]{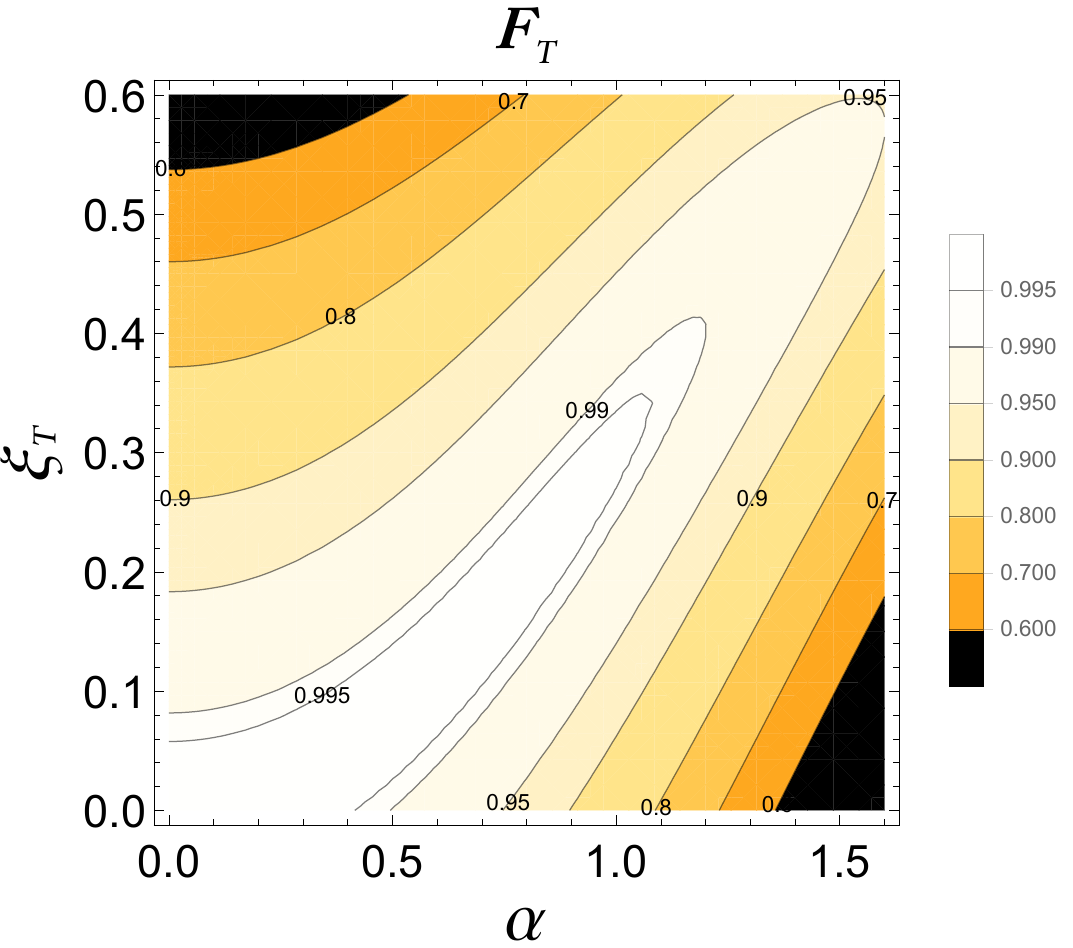}
\caption{Contour plot for fidelity $F_T$ between $\ket{Out_{\ket{1}} (\xi_T)}_A $ and $\ket{SCS^-_{i \alpha}}$ with respect to $\xi_T = T^2 \tanh \xi$ and $\alpha$.}
\label{fig:03}
\end{figure}

\subsection{Single-photon detection with a $BS^T$}
In the second step of the protocol, the action of the BS with transmission $T$ is described by the operator \cite{MSKim08} 
\begin{eqnarray}
\label{BS_T01}
BS^{T}_{A,A'} = \exp \left[ {\theta \over 2} \left( \hat{a}^{\dag} \hat{a'} -  \hat{a} \hat{a'}^{\dag} \right)\right].
\end{eqnarray}
In general, the transmission rate $T = \cos (\theta/2)$ in a mirowave BS is controlled by the common length of two transmission lines and a relatively short length is required for $T \approx 1$.

A single-photon detector is of essence for photonic QI processing and has been very recently demonstrated in experimental circuit-QED \cite{1microwaveD}. Because the energy of microwaves is in general some orders of amplitude weaker than optical photons and their wavelength are about the order of a chip size (milimeters to centimeters), it is very challenging for the detection of propagating microwave photons but a technical breakthrough has been developed to build a single microwave detector with a high efficiency and a low dark-count probability. In Ref.~\cite{1microwaveD}, four levels of a flux qubit are driven to be in dressed states to absorb a microwave photon in a half-wavelength resonator. Then, the readout of the qubit excited state is performed in a parametric phase-locked oscillator with detection efficiency above 0.6.

At the final part in Figure \ref{fig:01}, if a single photon is detected in mode $A'$, the outcome state in mode $A$ is given by
\begin{eqnarray}
\label{BS_T02}
\ket{Out_{\ket{1}} (\xi_T)}_A &=& {}_{A'} \bra{1} BS^{T}_{A,A'} \ket{Sq (\xi)}_A \ket{0}_{A'} \\
&=& N^T_{\xi} \left( \hat{a} \, T^{\hat{a}^\dag \hat{a}} \right) \ket{Sq (\xi)}_{A}, \nonumber
\end{eqnarray}
where $N^T_{\xi} = \left(1- (\xi_T )^2 \right)^{3/4} / \left( \sqrt{{\rm sech} \xi} \, \xi_T \right) $ for $\xi_T = T^2 \tanh \xi$. Note that ${}_{A'} \bra{1} BS^{T}_{A,A'} \ket{0}_{A'} \propto \hat{a} \, T^{\hat{a}^\dag \hat{a}} $ and $BS^{T}_{A,A'}$ can be decomposed into three terms, which are transformed to its normalization factor, $T^{\hat{a}^\dag \hat{a}}$, and $\hat{a}$ (see detail calculations in \cite{MSKim08}).
It shows that the projection to a single-photon state from a vacuum in mode $A'$ has the consequence of subtracting a photon with a BS from the other state in mode $A$ but to leave only the operator $\hat{a}$.
The fidelity $F_T$ between $\ket{SCS^-_{i \alpha} }$ and $\ket{Out_{\ket{1}} (\xi_T) } $ is given by
\begin{eqnarray}
\label{Fidelity02}
 F_{T} &&= |\bra{Out_{\ket{1}} (\xi_T) }  SCS^-_{i \alpha} \rangle |^2, \nonumber \\
&& = { |\alpha^2| \left(1- (\xi_T )^2 \right)^{3/2} \over \sinh (|\alpha|^2) } \exp \Big[  |\alpha^2| \,\xi_T \Big].
\end{eqnarray}
Figure \ref{fig:03} implies that the SCS $\ket{SCS^-_{i \alpha}}$ can be always found with the same fidelity $F_T$ along the contour line and that the appropriate parameter $\xi$ is fixed at a specific transmission rate $T$ to acquire the maximum $F_T$. 
For example, if a big squeezed vacuum is given with $\xi \gg 1$, its photon distribution in Fock states is very flat and $\xi_T$ is approximately $T^2$ due to $\tanh \xi\approx1$. Thus, a BS with $T\approx 1$ always gives us lower fidelity to the odd cat distribution while a BS with $T<0.6$ only does an opportunity to obtain good fidelity as shown in Fig.~\ref{fig:03}.

\begin{figure}[b]
\includegraphics[width=8.5cm,trim= 0cm 0cm 0cm 0cm]{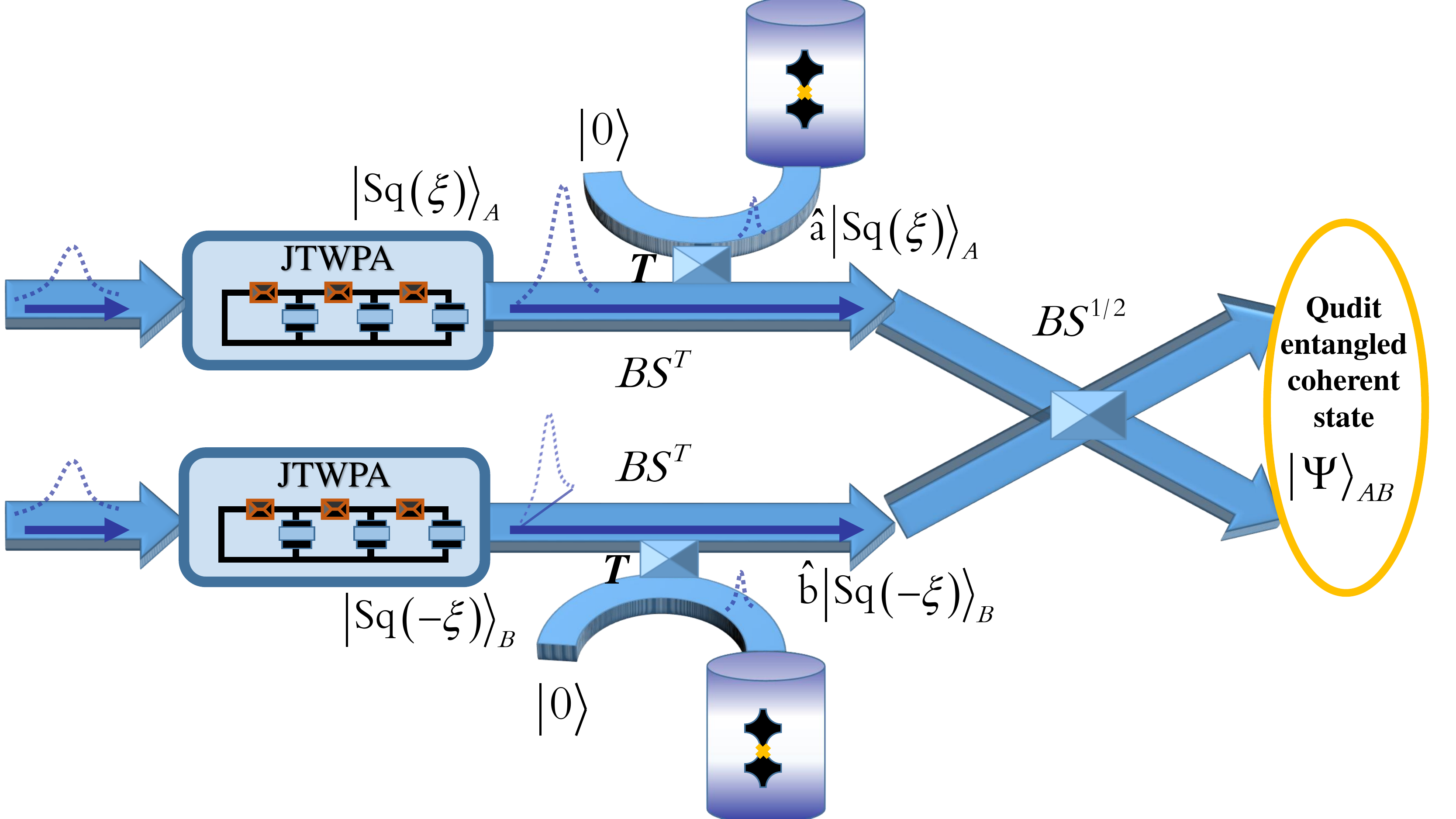}
\caption{Extended schematics for generating a traveling qudit ECS for $d=4$ in circuit-QED. The last $BS^{1/2}$ makes entanglement between two modes.}
\label{fig:04}
\end{figure}

Finally, homodyne (or heterodyne) measurement is a well established scheme for gaining the quadrature distribution in quantum optics. The quantum state tomography from this measurement scheme has been successfully performed in circuit-QED by a JPA squeezing method \cite{Lehnert11}. For example, an ordinary microwave amplifier (like a high-electron-mobility transistor) amplifies the input signal with large noise and the output states are too much blurred (or noisy) for quadrature detection. However, JPA can amplify the signal well enough with a phase-sensitive method. In particular, the output signal through JPA can clearly provide improvement of homodyne measurement to perform quantum tomography and even a realistic model with loss/noise conditions is demonstrated with an extra BS as an imperfect homodyne measurement in circuit-QED  \cite{Lehnert11}.

\section{Traveling qudit ECS}
We here propose a new scheme for building generalized entangled coherent states (ECSs) as a traveling entangled state in a representation of several coherent states in microwaves. The ECSs have been investigated for applications of QI processing \cite{Sanders92, Sanders12,JooPRL11} and the implementation schemes have been developed in quantum optics and circuit-QED \cite{GrangierECS, Yale-entangle}. An optical SVS is treated as an input even SCS in quantum optics and two SVSs are mixed with a BS to build a qubit ECS while the ECS in circuit-QED is trapped inside two microwave cavities. However, our target state is a more generalized ECS (named qudit ECS) and this qudit-type states are demonstrated and very useful for encoding quantum error-correcting codes under photon-loss \cite{qcMAP13}. 

The qudit ECS $\ket{\Psi_d}$ is defined as the entangled state with multiple coherent states. For example, for $d=4$,
\begin{eqnarray}
 \ket{\Psi_4}_{AB} &=& \sum_{i=0}^{3} c_i \ket{\tilde{i}_4}_A \ket{\tilde{i}_4}_B,
 \label{Sec4-01}
\end{eqnarray}
where $\ket{\tilde{0}_4} = \ket{\alpha}$, $\ket{\tilde{1}_4} = \ket{-i \alpha}$, $\ket{\tilde{2}_4} = \ket{-\alpha}$, and $\ket{\tilde{3}_4} = \ket{i \alpha}$ for coefficient $c_i$ \cite{Kim05}. 
Based on the above scheme of generating odd SCSs, it is assumed that each single photon-subtraction operation is implemented in modes $A$ and $B$ and $\ket{Out_{\ket{1}} (-\xi_T)}$ is approximately equal to $\ket{SCS^-_{\alpha}}$ in mode $B$ because $-\xi$ makes positive numbers in the terms of the power $l$ in Eq.~(\ref{Sq_V01}).
Once the single-photon detection occurs from the input SVSs simultaneously, the output state after a typical BS with $T=1/2$ is very close to the qudit ECS such as
\begin{eqnarray}
&& \ket{\tilde{\Psi}_4}_{AB} = BS^{1/2}_{AB} \, \ket{SCS^-_{i \alpha}}_A  \, \ket{SCS^-_{\alpha}}_B ,  \label{Sec4-02} \\
&& ~\propto\ket{\tilde{0}_4}_A \ket{\tilde{0}_4}_B - \ket{\tilde{3}_4}_A \ket{\tilde{1}_4}_B  - \ket{\tilde{1}_4}_A \ket{\tilde{3}_4}_B + \ket{\tilde{2}_4}_A \ket{\tilde{2}_4}_B, \nonumber
\end{eqnarray}
where  
$
BS^{1/2}_{AB} \, \ket{i \alpha}_A  \, \ket{SCS^-_{\alpha}}_B 
~\propto~  \ket{\alpha}_A \ket{\alpha}_B - \ket{-i\alpha}_A \ket{i \alpha}_B
$ and
$
BS^{1/2}_{AB} \, \ket{-i \alpha}_A  \, \ket{SCS^-_{\alpha}}_B 
~\propto~  \ket{i\alpha}_A \ket{-i\alpha}_B - \ket{-\alpha}_A \ket{-\alpha}_B
$ upto linear phase operations.

From the qudit ECS $\ket{\tilde{\Psi}_4}_{AB}$, a microwave detection in mode $A$ (or $B$) possibly generates a superposition of four coherent states which could be a carrier of error-correctable logical qubits against photon losses \cite{qcMAP13}. However, because the amplitude of these coherent states are not large enough for fault-tolerant QI processing, one may need to perform the amplification schemes of traveling coherent states (or SCSs) to maintain the orthogonality among the coherent states \cite{amplification}. 

As shown in Fig. \ref{fig:04}, the final output starting from two SVSs is explicitly given by
\begin{eqnarray}
\ket{\Psi}_{AB}& &= BS^{1/2}_{AB} \, \ket{Out_{\ket{1}} (\xi_T)}_A \ket{Out_{\ket{1}} (-\xi_T)}_B, 
\label{Sec4-04} \\
 & & =  \sum_{n=2}^{\infty} {\tau_n}\Big(  \ket{n-2}_A \ket{n}_B - \ket{n}_A \ket{n-2}_B \Big), \nonumber 
\end{eqnarray}
where $\tau_n = {\rm sech} \,\xi \,(N^T_{\xi})^2 (- \xi_T )^n $. 
 
Therefore, $\ket{\Psi}_{AB} \approx \ket{\tilde{\Psi}_4}_{AB}$ due to $ \ket{Out_{\ket{1}} (\xi_T)} \approx \ket{SCS^-_{i \alpha}}$. For example, if a two-photon state $\ket{2}_A$ is measured in $\ket{\Psi}_{AB}$, the outcome is given by $\ket{0^L} \propto \tau_4 \ket{4}_B -\tau_2 \ket{0}_B$. On the other hand, we can obtain $\ket{1^L} \propto \tau_6 \ket{6}_B -\tau_4 \ket{2}_B$ with detection of $\ket{4}_A$ \cite{catcode}. Thus, a traveling logical qubit can be built in the superposition of only even-photon states and is error correctable by parity measurements \cite{Yale-QECC}.

Interestingly, the state $\ket{\Psi}_{AB} $ in Eq.~(\ref{Sec4-04}) is also known as a robust quantum resource for quantum sensing against photon losses \cite{Dowling}. The N00N state, given by $(\ket{N}_A \ket{0}_B + \ket{0}_A \ket{N}_B)/\sqrt{2}$ for photon number $N$, provides the Heisenberg limit for quantum phase metrology but is fragile against photon losses while $\ket{\Psi}_{AB} $ still keep the phase information in the presence of channel losses. For example, if photon losses occur in mode $B$, the N00N state loses the superposition of two-mode states and becomes a mixed state only in mode $B$ without phase information while other resilient states (like $\ket{\Psi}_{AB}$ and $\ket{\tilde{\Psi}_4}_{AB}$) can be two-mode mixed states with many Fock-state components, which contain imprinted phase information \cite{Ian09}.
 
\section{Conclusion and remarks}
In summary, we propose an implementation scheme of traveling SCSs and qudit ECS in circuit-QED. A single-photon subtraction method is currently feasible using newly developed microwave-photon detectors and a nonclassical state (i.e., SVS) is manifested as an input resource state to generate approximated SCSs. The fidelity of the output states with the target odd SCSs reaches to above 0.95 for $0 \le \alpha \le 1.6$ theoretically. Furthermore, if one can perform the scheme of the odd SCS twice simultaneously, a traveling qudit ECS could be feasible. 

A few obstacles still remain for demonstrating our scheme in the state-of-the-art techniques of circuit-QED. The amount of losses are currently inevitable in traveling microwaves along transmission lines. The current schemes of a single-microwave detector include a few circulators in which microwave photons suffer losses, so our setup might be built in a single superconductor chip without a circulator. The single microwave detectors in \cite{1microwaveD} are equivalent to on-off detectors and only can be used for our scheme with $T\approx 1$ due to maintaining a very low average photon number in the detection line. This causes a low success probability to generate the odd SCS. Thus, the improvement of the single-photon detector will provide higher success probability for the scheme.

\section{Acknowledgement}
This work was supported by the KIST Institutional Program (Project No. 2E26680-16-P025). We thank S. Kono, E. Ginossar, Y. Nakamura, and H. Paik
for useful discussions.

\end{document}